\documentclass[fleqn,usenatbib]{mnras}

\usepackage{newtxtext,newtxmath}

\usepackage[T1]{fontenc}

\DeclareRobustCommand{\VAN}[3]{#2}
\let\VANthebibliography\thebibliography
\def\thebibliography{\DeclareRobustCommand{\VAN}[3]{##3}\VANthebibliography}

\usepackage{graphicx}	
\usepackage{amsmath}	
\usepackage{subcaption}

\title[Timing Analysis of Swift J1727.8--1613]{Timing analysis of the newly discovered black hole candidate Swift J1727.8--1613 with \emph{Insight}-HXMT }

\author[Yu et al.]{
Wei Yu,$^{1,2}$
Qing-Cui Bu,$^{3}$\thanks{E-mail: bu@astro.uni-tuebingen.de}
Shuang-Nan Zhang,$^{1,2}$
He-Xin Liu,$^{1}$
Liang Zhang,$^{1}$
Lorenzo Ducci,$^{3,4}$
\newauthor
Lian Tao,$^{1}$
Andrea Santangelo,$^{3}$
Victor Doroshenko,$^{3}$
Yue Huang,$^{1}$
Zi-Xu Yang,$^{5}$
and Jin-Lu Qu,$^{1,2}$
\\
$^{1}$Key Laboratory of Particle Astrophysics, Institute of High Energy Physics, Chinese Academy of Sciences, 19B Yuquan Road, Beijing 100049, China\\
$^{2}$University of Chinese Academy of Sciences, Chinese Academy of Sciences, Beijing 100049, China\\
$^{3}$Institut f{\"u}r Astronomie und Astrophysik, Kepler Center for Astro and Particle Physics, Eberhard Karls Universit{\"a}t, 72076 T{\"u}bingen, Germany\\
$^{4}$ISDC Data Center for Astrophysics, Universit\'e de Gen\`eve, 16 chemin d'\'Ecogia, 1290 Versoix, Switzerland\\
$^{5}$School of Physics and Optoelectronic Engineering, Shandong University of Technology, Zibo 255000, China
}
\date{Accepted XXX. Received YYY; in original form ZZZ}

\pubyear{2023}

\begin{document}
\label{firstpage}
\pagerange{\pageref{firstpage}--\pageref{lastpage}}
\maketitle

\begin{abstract}
We present the results obtained from an X-ray timing study of the new black hole candidate (BHC) Swift J1727.8--1613. The work is based on Hard X-ray Modulation Telescope (\emph{Insight}-HXMT) observations carried out during the 2023 outburst.
Prominent type-C low-frequency Quasi-periodic Oscillations (LFQPOs) are detected throughout the observations. With the substantial effective area of the Insight-HXMT at high energies, we examine the energy dependence of various parameters, including the centroid frequency, fractional rms, and phase lags of the type-C QPOs.
Our findings align closely with those observed in high-inclination systems. During the initial stage of the outburst, a peaked noise component is also detected, the frequency of which is highly correlated with the LFQPO frequency, aligning with the Psaltis-Belloni-van der Klis (PBK) relation. By assuming that the peaked noise originates from the precession of the accretion disc, the spin of this source can be constrained. Our results suggest that this source may possess a high spin.
\end{abstract}




\section{Introduction}

Black hole low-mass X-ray binaries (BH-LMXB) predominantly consist of transient systems wherein a black hole accretes matter from its companion star through an accretion disc \citep{shakura1973black}. These transients, termed black hole transients (BHTs), largely remain in a quiescent state, punctuated by occasional outbursts that persist for several weeks to months. Such outbursts are posited to arise from the system's inherent instabilities \citep{cannizzo1995accretion, lasota2001disc}. During these outbursts, the source luminosity can approach the Eddington limit. Concurrently, there is a marked shift in both the energy spectral properties and rapid variability, facilitating the categorization into distinct spectral states \citep{belloni2010states}.

A consistent pattern of X-ray spectral evolution in BHTs has been observed across most systems during an outburst, commonly referred to as the hardness-intensity diagram (HID). Initially, as the system emerges from its quiescent state, it transitions into the low hard state (LHS). In the LHS, the X-ray emission predominantly originates from non-thermal coronal photons. These photons are believed to result from the inverse Compton scattering between soft disc photons and hot electrons present in the corona. This state's X-ray spectrum is best characterized by a phenomenological power-law featuring a high energy cutoff \citep{zdziarski2004radiative, remillard2006x, done2007modelling}. Within the power density spectrum (PDS), one can detect strong band-limited noise and low-frequency quasi-periodic oscillations (LFQPOs). As luminosity increases, the source transitions to the high soft state (HSS), where thermal disc emission dominates the spectrum. The X-ray spectrum in the HSS agrees well with a multi-temperature disc-blackbody component \citep{remillard2006x, you2016testing}, and the corresponding PDS manifests as a power-law-shaped red noise. Between the hard and soft states, two intermediate states, namely the hard intermediate state (HIMS) and the soft intermediate state (SIMS), are defined based on their distinct spectra and timing properties. \citep{homan2005evolution, belloni2005evolution}. Such transitions typically coincide with changes in the LFQPO types: type-C QPOs are mainly present in the HIMS, while type-B and type-A QPOs are exclusive to the SIMS and HSS \citep{2005ApJ...629..403C,2014MNRAS.443.3270M,2019NewAR..8501524I}. 

The study of LFQPOs is essential to our understanding of the accretion flow around black holes, though their origin is still controversial. Several models have been proposed to explain the dynamical origin of LFQPOs considering either the instability or a geometric effect of the accretion flow, among which the most promising model is the Lense-Thirring (L-T) precession model that assumes that LFQPOs are generated by the relativistic precession of an inner hot accretion flow \citep{ingram2009low, you2018x, you2020x} or a small-scale jet \citep{2021NatAs...5...94M,2023ApJ...948..116M}. The study of the energy-dependent timing properties of QPOs, such as the fractional rms amplitude, centroid frequency, and the lags between different
energy bands can help us understand the radiative process behind QPOs.

The new X-ray transient, Swift J1727.8-1613, was discovered on August 24, 2023 \citep{2023GCN.34536....1L,2023GCN.34537....1P}. This source was initially recognized as GRB 230824A. However, the subsequent observation revealed that the source exhibited a rapid flux increase and was identified as a new galactic X-ray transient \citep{2023ATel16205....1N,2023ATel16206....1N}. Further optical \citep{2023ATel16208....1C}, X-ray \citep{2023ATel16207....1O}, and radio \citep{2023ATel16211....1M} observations indicate the source as a black hole candidate. A distance to the source of $2.7\pm0.3$ kpc was measured by \citet{2024A&A...682L...1M}. LFQPOs have been detected by \emph{Swift}/XRT, \emph{NICER}, \emph{AstroSat}/LAXPC, and \emph{INTEGRAL} \citep{2023ATel16215....1P,2023ATel16219....1D,2023ATel16235....1K,2023arXiv231006697M}. The polarized X-ray signals have been detected by \emph{IXPE} \citep{2023ApJ...958L..16V,2023arXiv231105497I}.

In this paper, we study the temporal variation of the source using \emph{Insight}-HXMT observations. In Section~\ref{sec2}, we describe \emph{Insight}-HXMT observations and data reduction methods. The results are presented in Section~\ref{sec3}. Discussions and Conclusions follow in Sections~\ref{sec4} and \ref{sec5}.

\section{DATA REDUCTION}
\label{sec2}

The Hard X-ray Modulation Telescope, known as \emph{Insight}-HXMT \citep{zhang2014introduction}, consists of three groups of instruments: the high-energy X-ray telescope \citep[HE: 20-250 keV;][]{2020SCPMA..6349503L}, the medium-energy X-ray telescope \citep[ME: 5-30 keV;][]{2020SCPMA..6349504C}, and the low-energy X-ray telescope \citep[LE: 1-15 keV;][]{2020SCPMA..6349505C}. There are three types of Field of View (FoV): 1$^\circ$ × 6$^\circ$ (i.e., the small FoV), 6$^\circ$ × 6$^\circ$ (i.e., the large FoV), and the blind FoV used to estimate the particle induced instrumental background. More details about \emph{Insight}–HXMT can be found in \citet{zhang2020overview}.
Following MAXI/GSC and Swift/BAT discovery of Swift J1727.8--1613, we triggered \emph{Insight}-HXMT Target of Opportunity (ToO) observations. Our follow-up observations started on August 25, 2023. However, due to the obscuration by the sun, we had to conclude the observation earlier than anticipated, on October 4, 2023.

The data are processed using the \texttt{hpipeline} under \emph{Insight}-HXMT Data Analysis Software (HXMTDAS) version 2.05. The data are filtered using the criteria recommended by the \emph{Insight}-HXMT team: (1) pointing offset
angle less than $0.04^\circ$; (2) Earth elevation angle larger than
$10^\circ$; (3) the vaule of the geomagnetic cutoff rigidity larger
than 8 GV; (4) at least 300 s before and after the passage of the South Atlantic Anomaly. To avoid potential contamination from bright earth and nearby sources, only the small field of view (FoVs) were applied.

\section{Results}
\label{sec3}
\subsection{Fundamental diagrams}

\begin{figure}
	\includegraphics[width=\columnwidth]{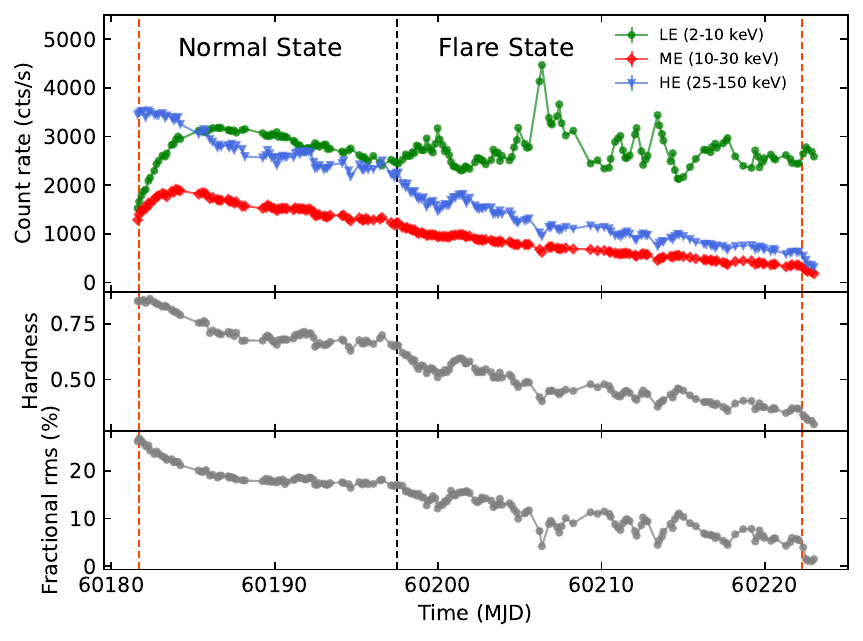}
    \caption{Light curves, hardness ratio and total fraction rms evolution for Swift J1727.8--1613 during its 2023 outburst observed by \emph{Insight}–HXMT. The top panel presents the light curves by three energy bands: 2–10 keV (LE, green points), 10–35 keV (ME, red points), and 25–150 keV (HE, blue points). The middle panel shows the hardness between 4–10 keV (LE) and 2–4 keV (LE) count rate. We show the total fraction rms of the LE light curve in the bottom panel. The two vertical orange dashed lines indicate the period in which the LFQPOs are detected.}
    \label{fig1}
\end{figure}

\begin{figure}
	\includegraphics[width=\columnwidth]{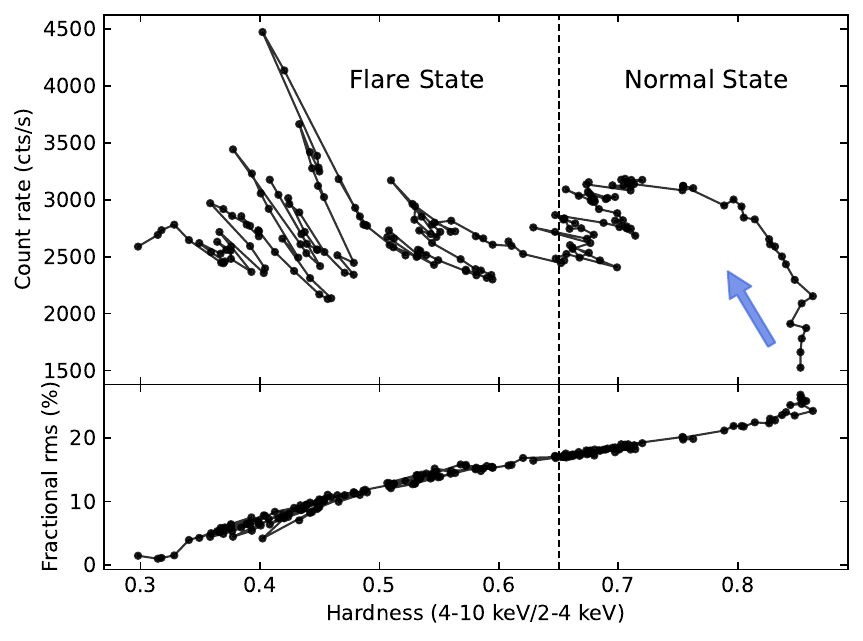}
    \caption{The HID and HRD of Swift J1727.8--1613 in this outburst are presented in the top and bottom panels respectively. Intensity is the LE count rate in the 2-10 keV. The hardness is defined as the ratio between LE’s 4–10 keV and 2–4 keV count rate.}
    \label{fig2}
\end{figure}

\begin{figure*}
    \includegraphics[width=17cm]{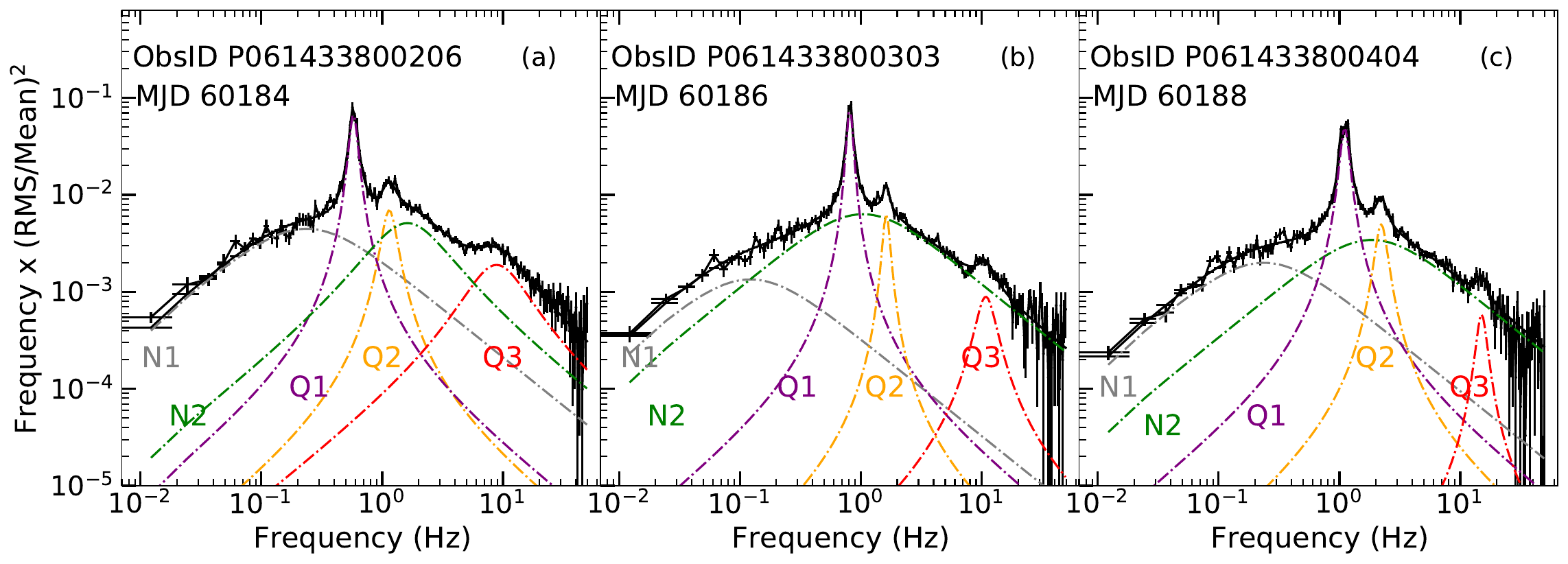}
    \includegraphics[width=17cm]{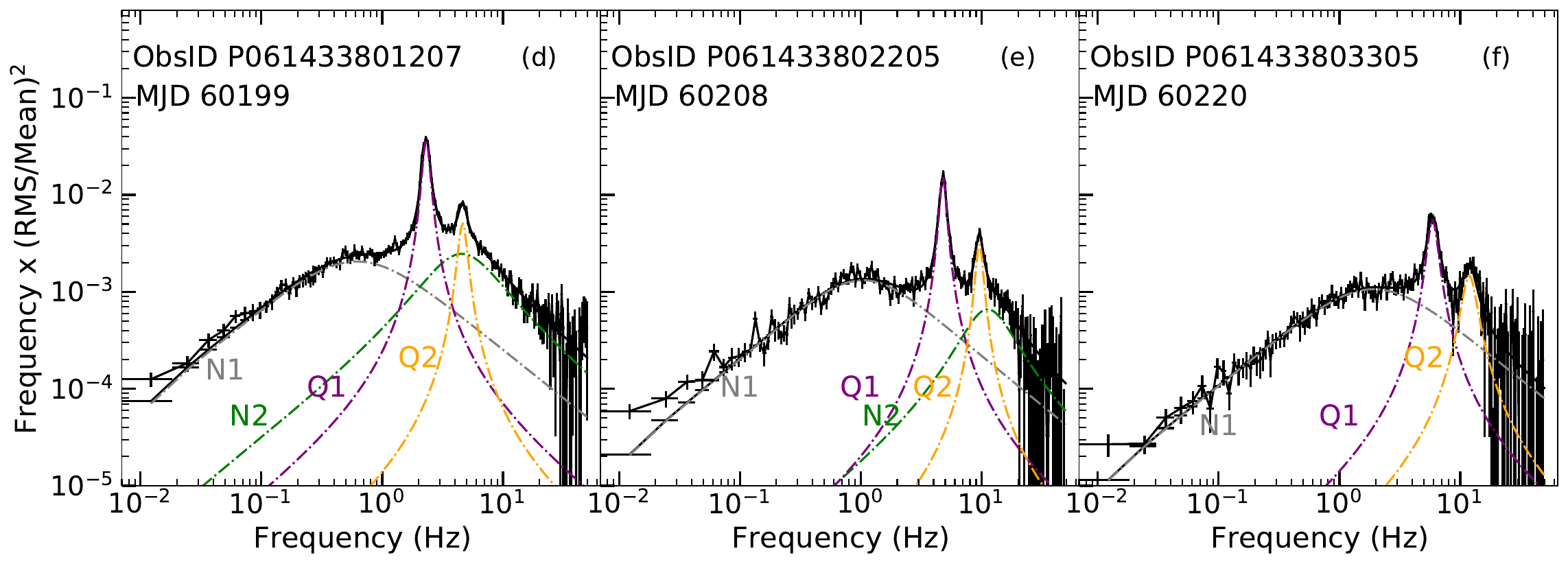}
    \caption{Six representative power spectra of the outburst of Swift J1727.8--1613. Dashed lines represent the best-fitting Lorentzians. Q1, Q2 and Q3 represent the QPO, its second harmonic and peaked noise, respectively. N1 and N2 represent the two broadband noise components on different timescales.}
    \label{fig3}
\end{figure*}

\begin{figure}
	\includegraphics[width=\columnwidth]{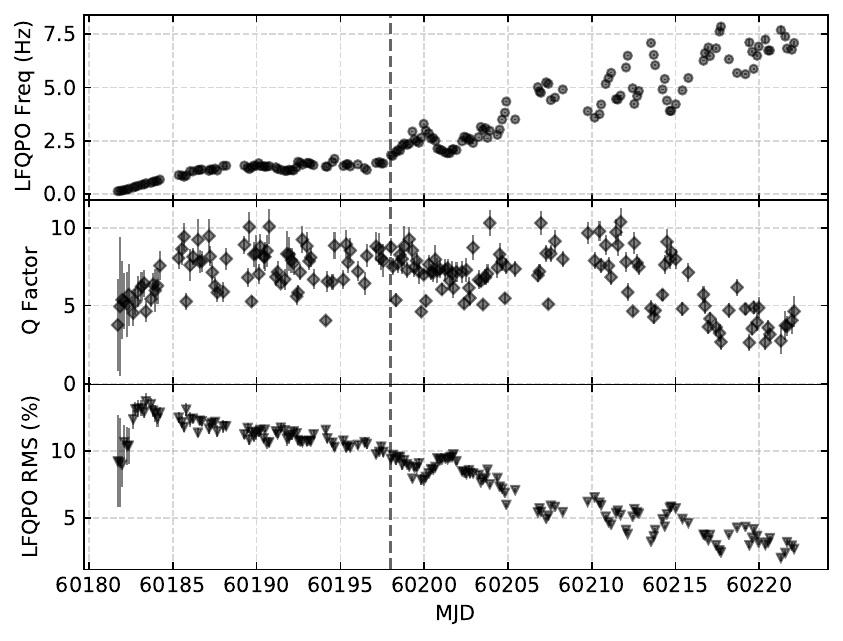}
    \caption{The LFQPO frequency, Q factor, and fractional rms from the LE energy band as functions of time for Swift J1727.8--1613. The vertical black dashed line represents the transition date from the `normal' state to the `flare' state.}
    \label{fig5}
\end{figure}

\begin{figure}
	\includegraphics[width=\columnwidth]{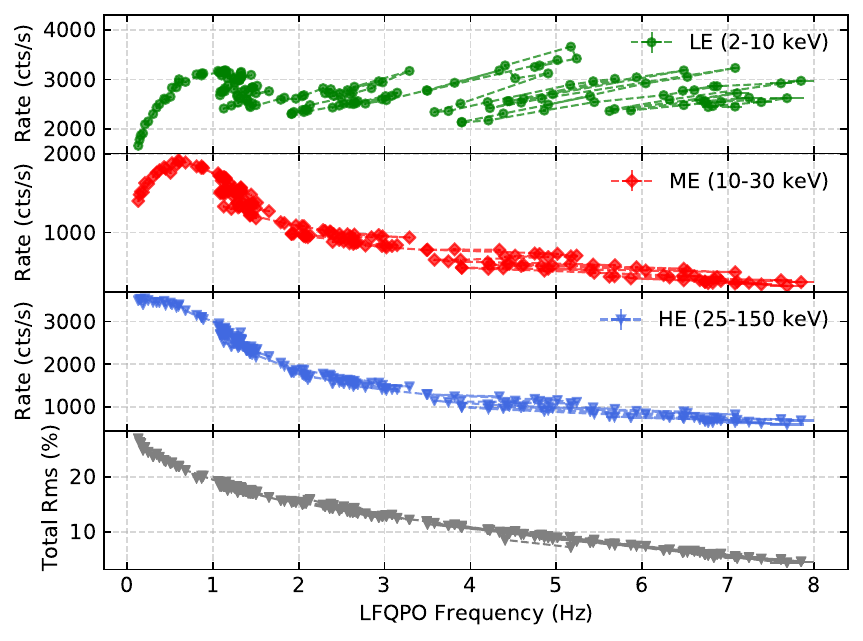}
    \caption{The LE, ME, HE rate and total fractional rms as functions of QPO frequency.}
    \label{fig6}
\end{figure}

\begin{figure}
	\includegraphics[width=\columnwidth]{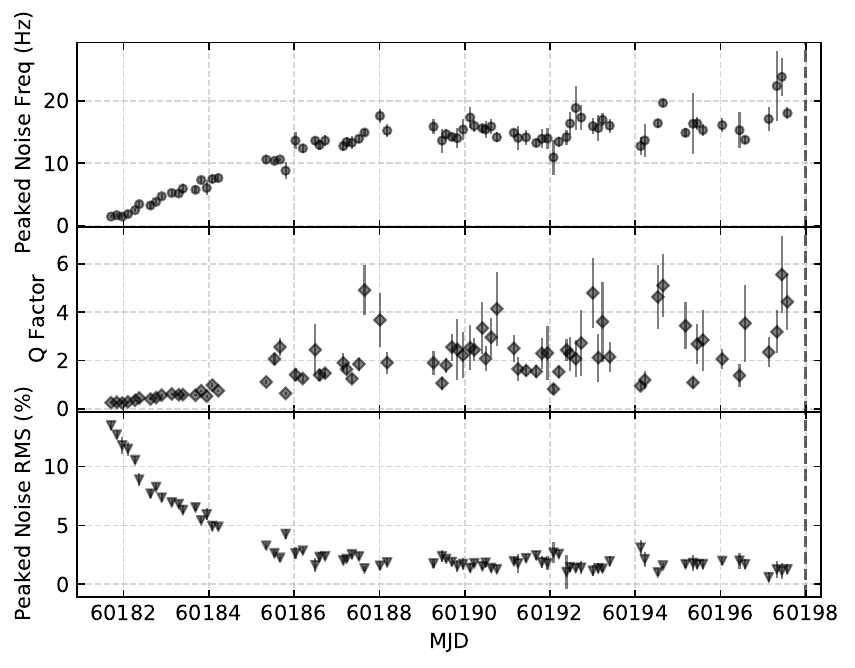}
    \caption{The peaked noise frequency, Q factor, and fractional rms in the LE energy band as functions of time for Swift J1727.8--1613. The vertical black dashed line represents the transition from the `normal' state to the `flare' state.}
    \label{fig7}
\end{figure}

\begin{figure*}
	\includegraphics[width=\columnwidth]{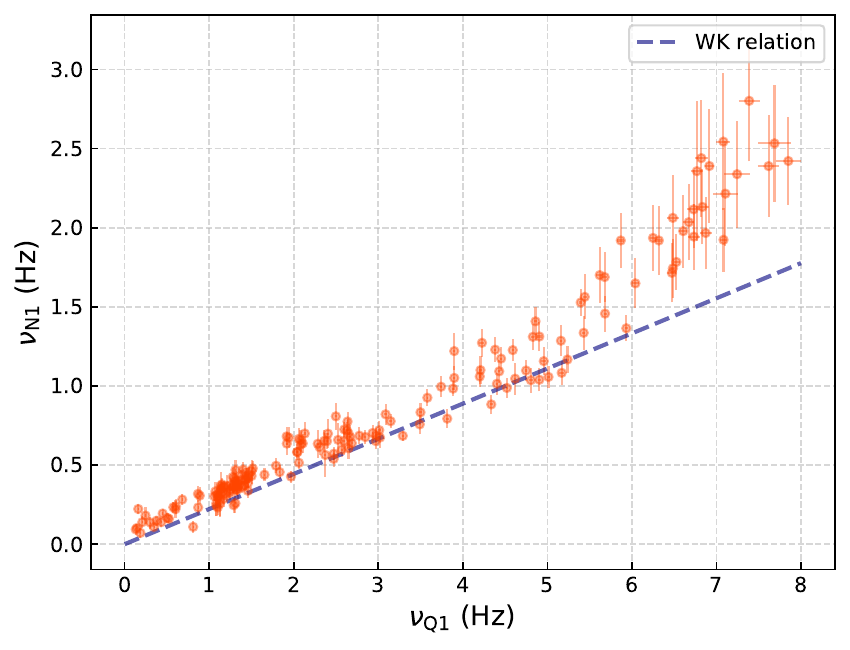}
        \includegraphics[width=\columnwidth]{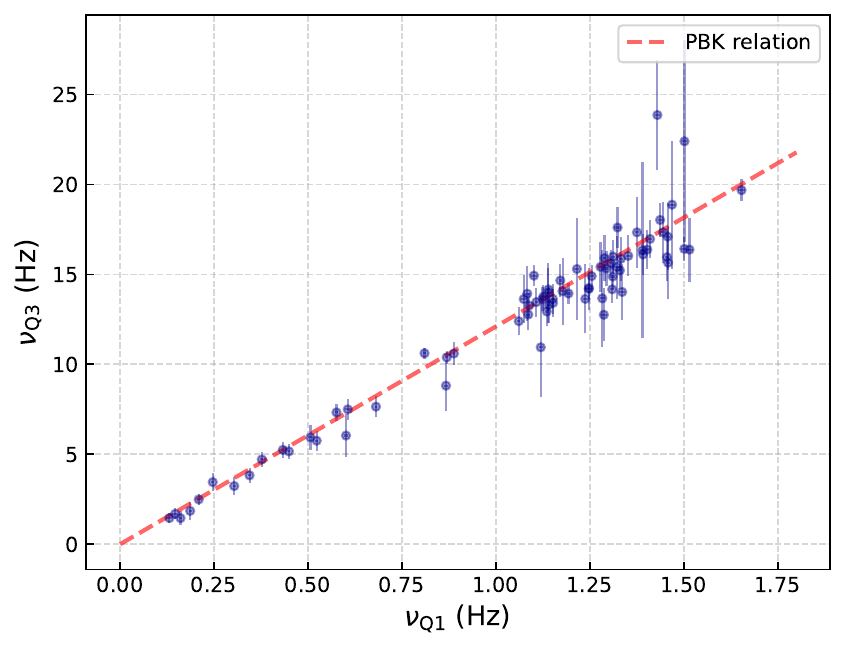}
    \caption{Relation of the frequencies of LFQPO (Q1) with low-frequency noise (N1) and peaked noise (Q3). The red dashed line represents the PBK correlation from \citet{1999ApJ...520..262P}. The blue dashed line represents the WK relation as shown in \citet{1999ApJ...514..939W}.}
    \label{fig8}
\end{figure*}

\begin{figure*}
	\includegraphics[width=5.7cm]{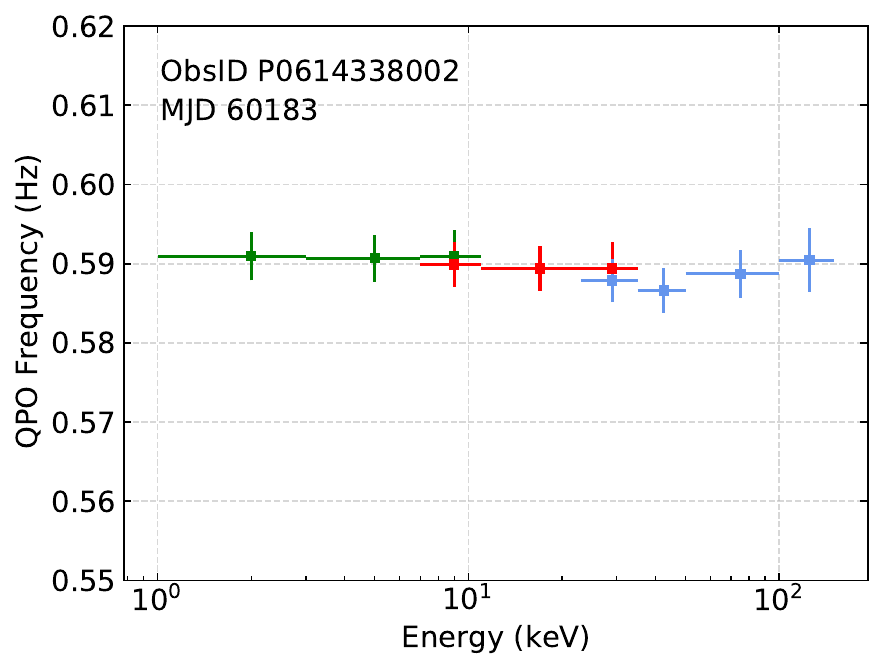}
        \includegraphics[width=5.7cm]{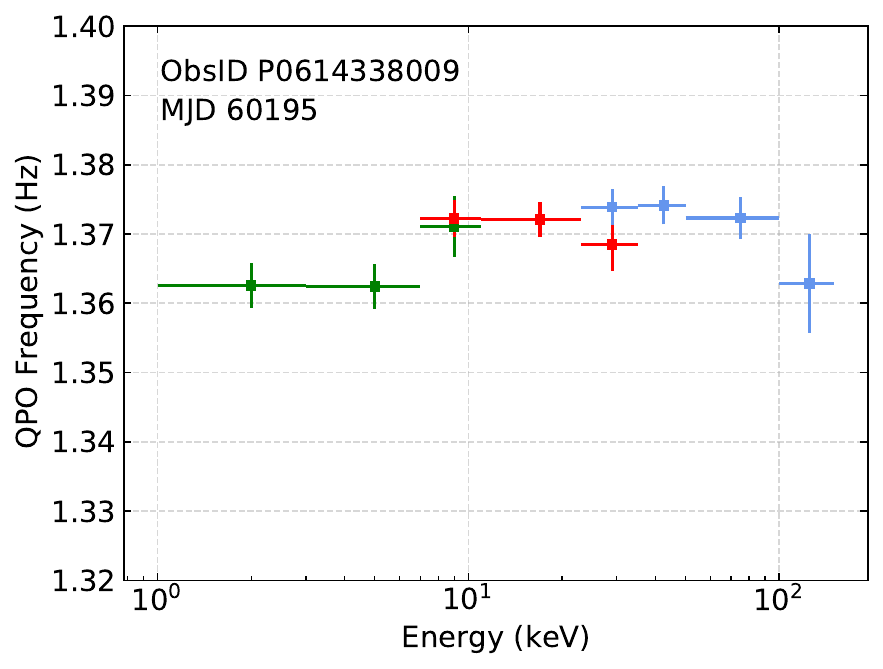}
        \includegraphics[width=5.7cm]{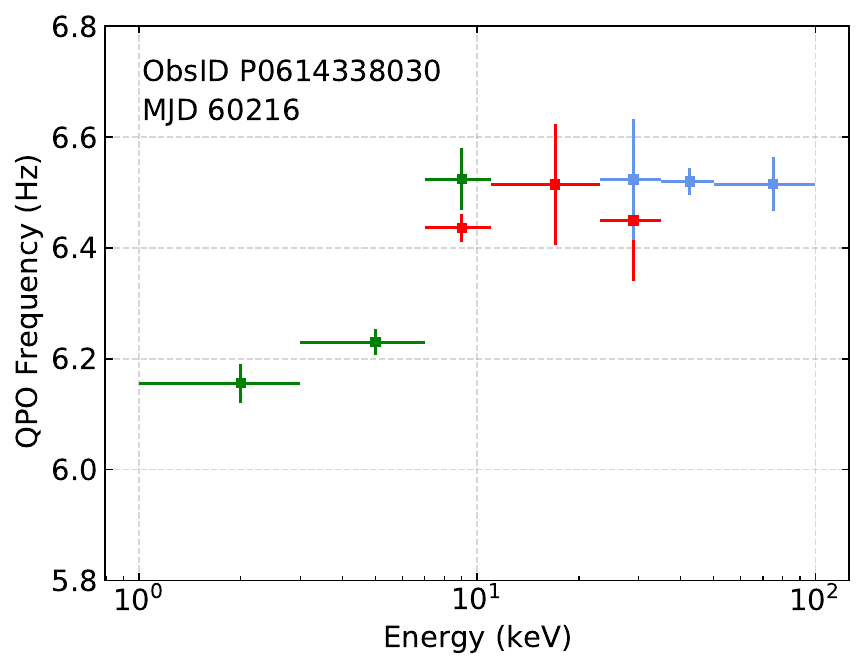}
        \includegraphics[width=5.7cm]{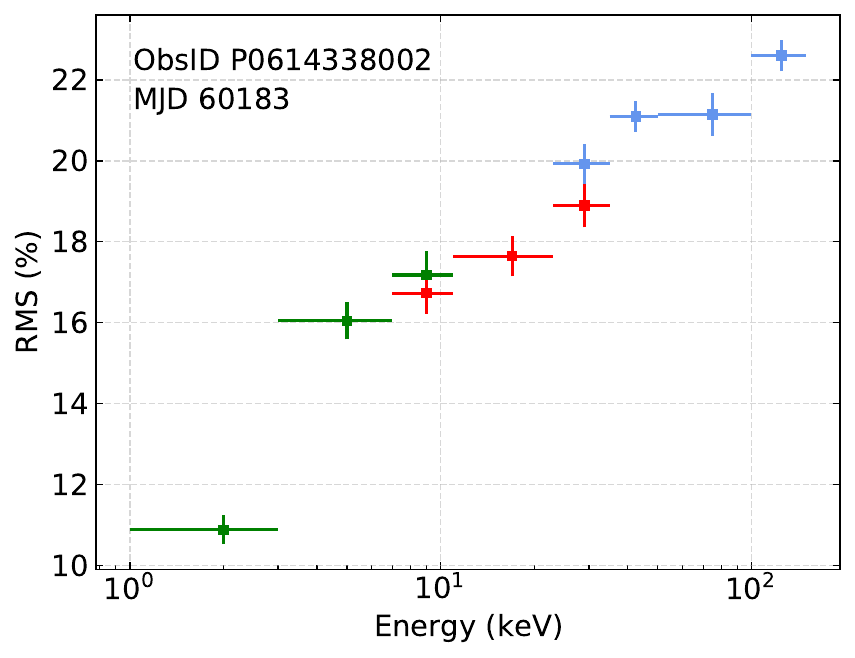}
        \includegraphics[width=5.7cm]{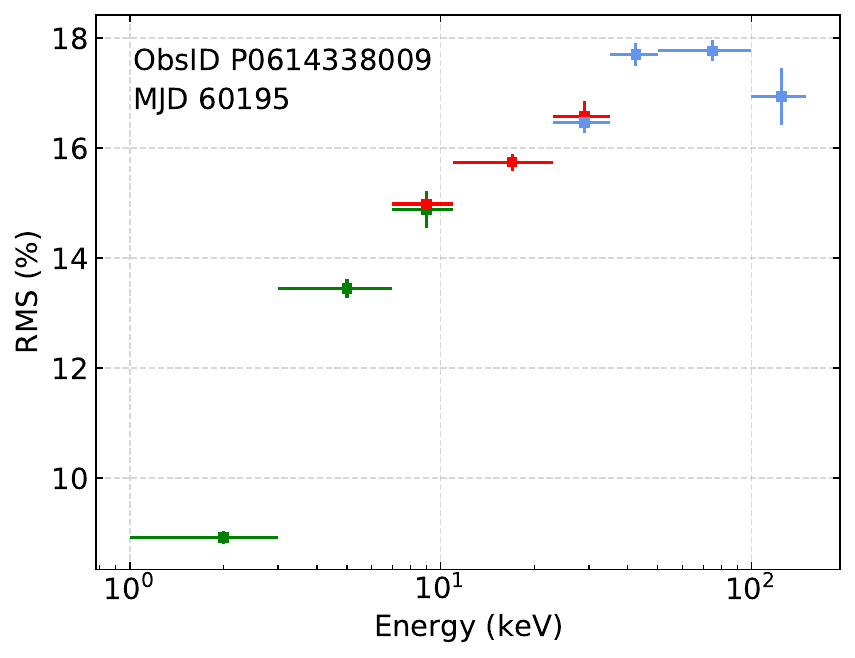}
        \includegraphics[width=5.7cm]{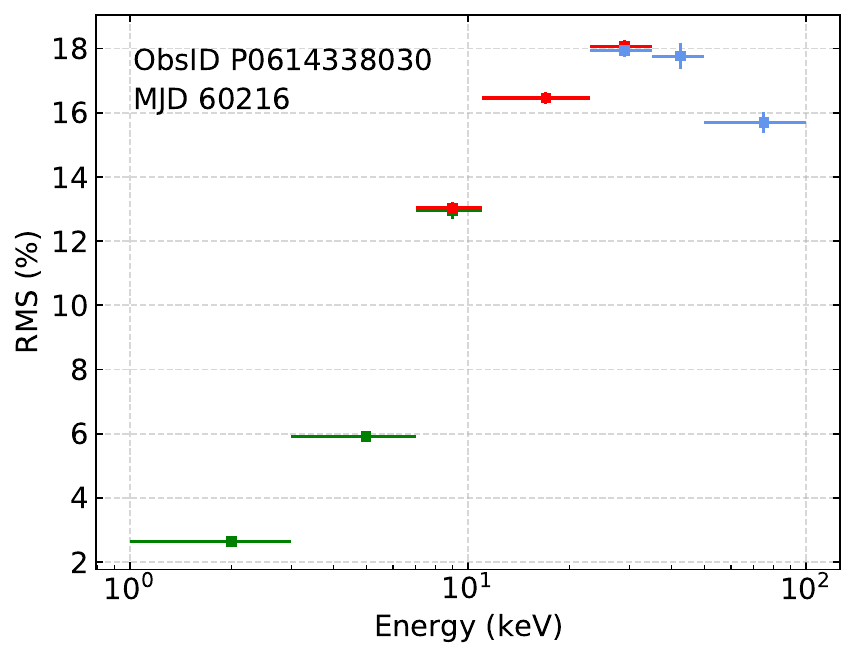}
    \caption{LFQPO centroid frequency and fractional rms amplitude as a function of energy for three typical observations. The green, red and blue points represent LE, ME and HE data respectively.}
    \label{fig9}
\end{figure*}

\begin{figure*}
    \begin{subfigure}{6cm}
        \centering
        \includegraphics[width=\linewidth]{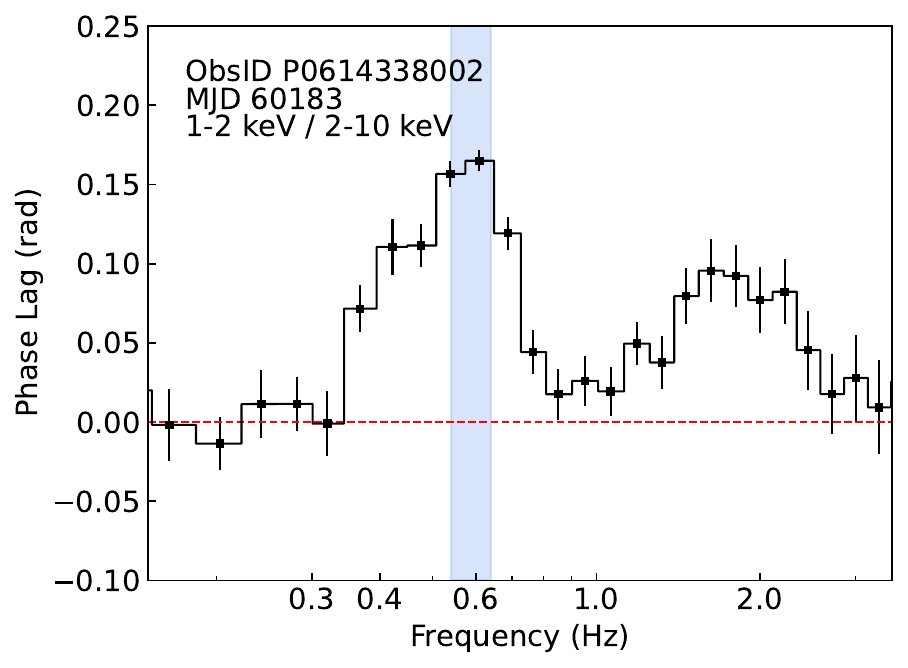}
        \caption{}
        \label{fig10:sub1}
    \end{subfigure}
    \hspace{1cm}
    \begin{subfigure}{6cm}
        \centering
        \includegraphics[width=\linewidth]{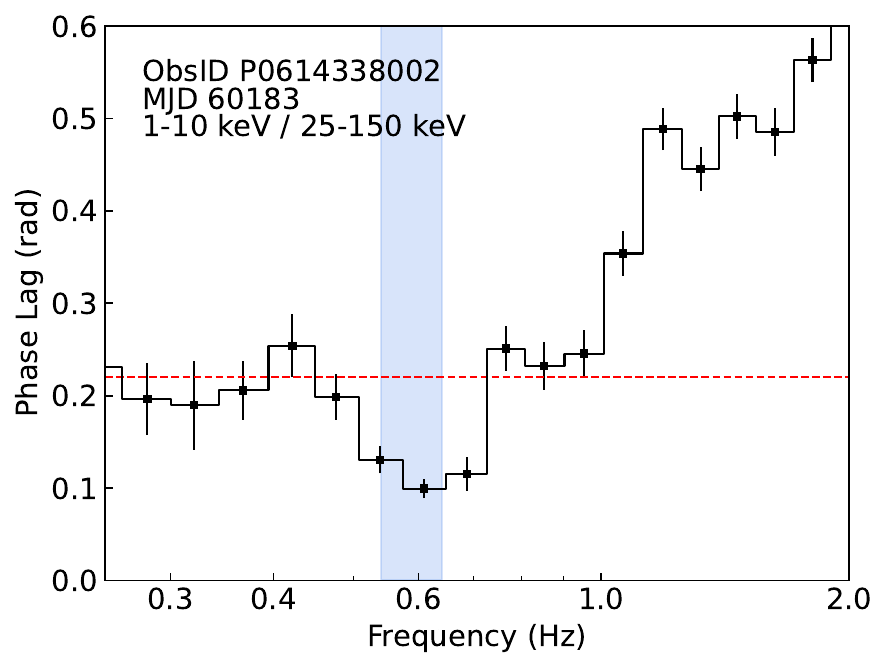}
        \caption{}
        \label{fig10:sub2}
    \end{subfigure}
    \begin{subfigure}{6cm}
        \centering
        \includegraphics[width=\linewidth]{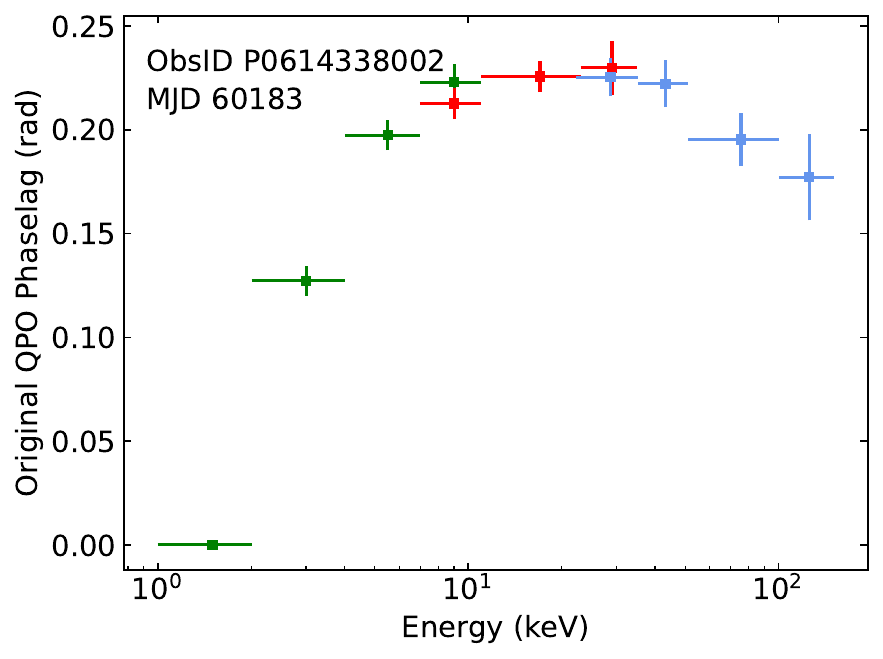}
        \caption{}
        \label{fig10:sub3}
    \end{subfigure}
    \hspace{1cm}
    \begin{subfigure}{6cm}
        \centering
        \includegraphics[width=\linewidth]{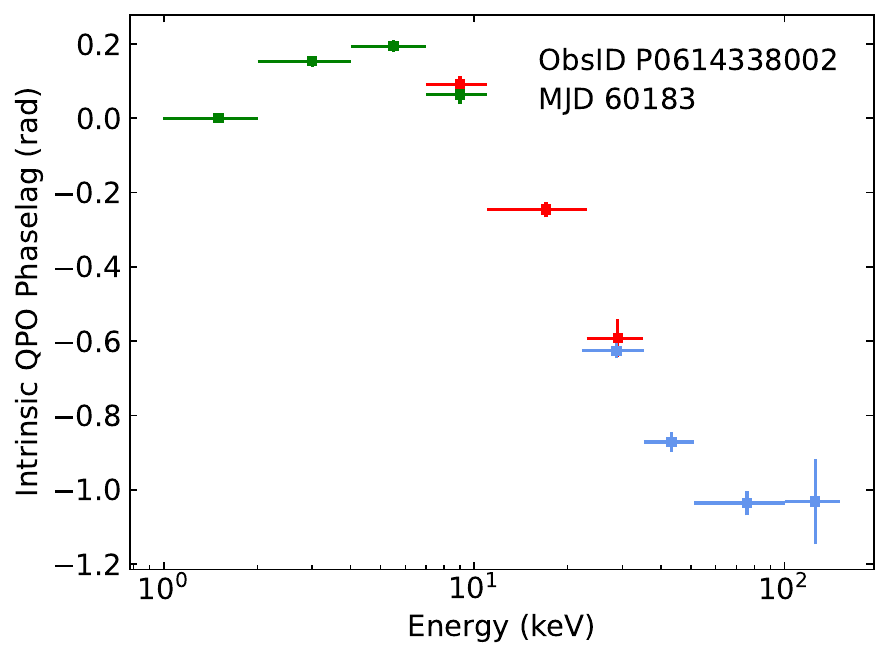}
        \caption{}
        \label{fig10:sub4}
    \end{subfigure}
    \caption{Phase lag spectra with a low (\textasciitilde0.6 Hz) type-C QPO frequency. The reference energy band used to calculate the lag-energy spectra is 1-2 keV. (a) Phase lags as a function of frequency between 1-2 keV and 2-10 keV. (b) Phase lags as a function of frequency between 1-10 keV and 25-150 keV. (c) The "original" LFQPO phase lags as a function of photon energy. (d) The "intrinsic" LFQPO phase lags as a function of photon energy. }
    \label{fig10}
\end{figure*}

\begin{figure*}
    \begin{subfigure}{6cm}
        \centering
        \includegraphics[width=\linewidth]{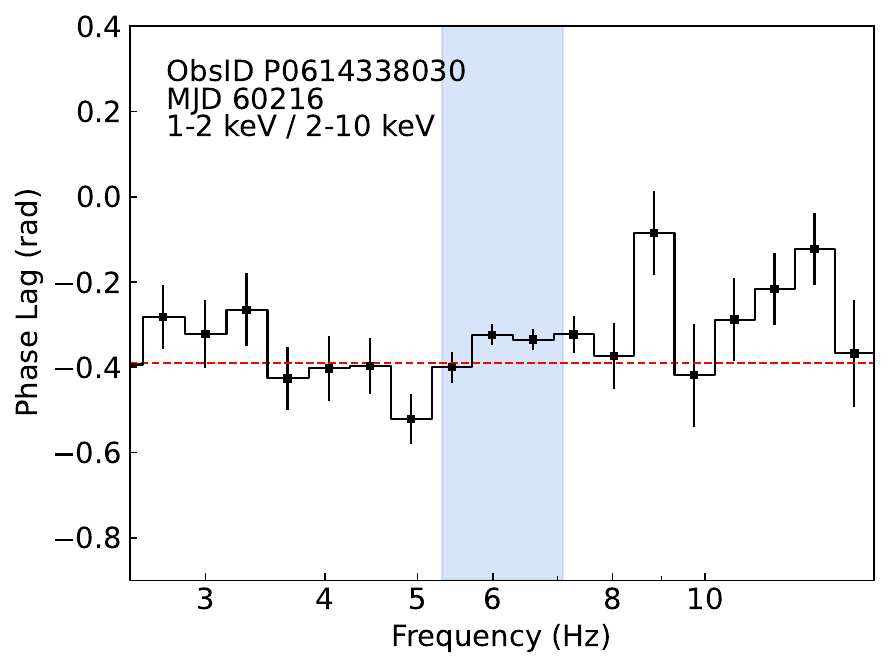}
        \caption{}
        \label{fig:sub1}
    \end{subfigure}
    \hspace{1cm}
    \begin{subfigure}{6cm}
        \centering
        \includegraphics[width=\linewidth]{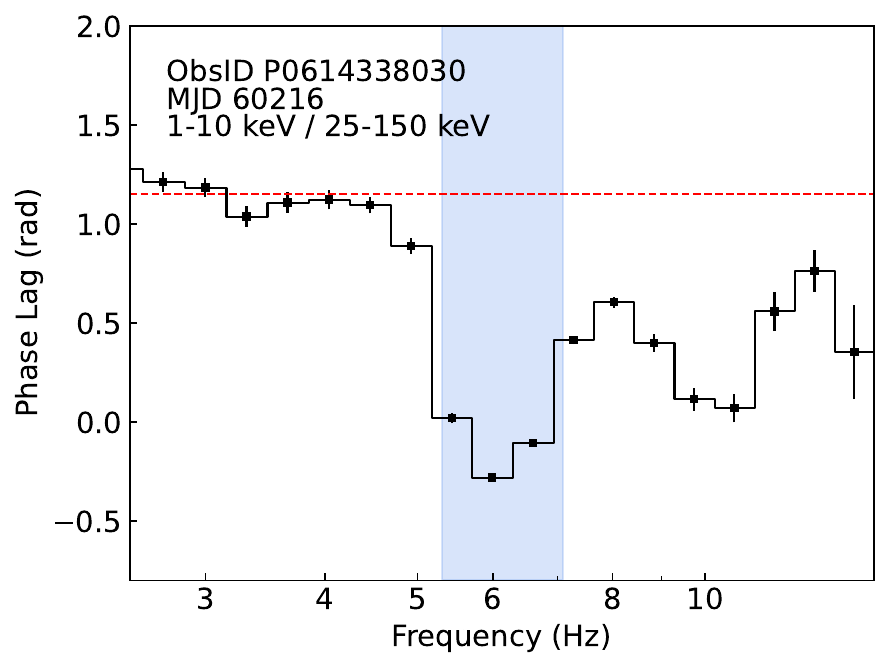}
        \caption{}
        \label{fig:sub2}
    \end{subfigure}
    \begin{subfigure}{6cm}
        \centering
        \includegraphics[width=\linewidth]{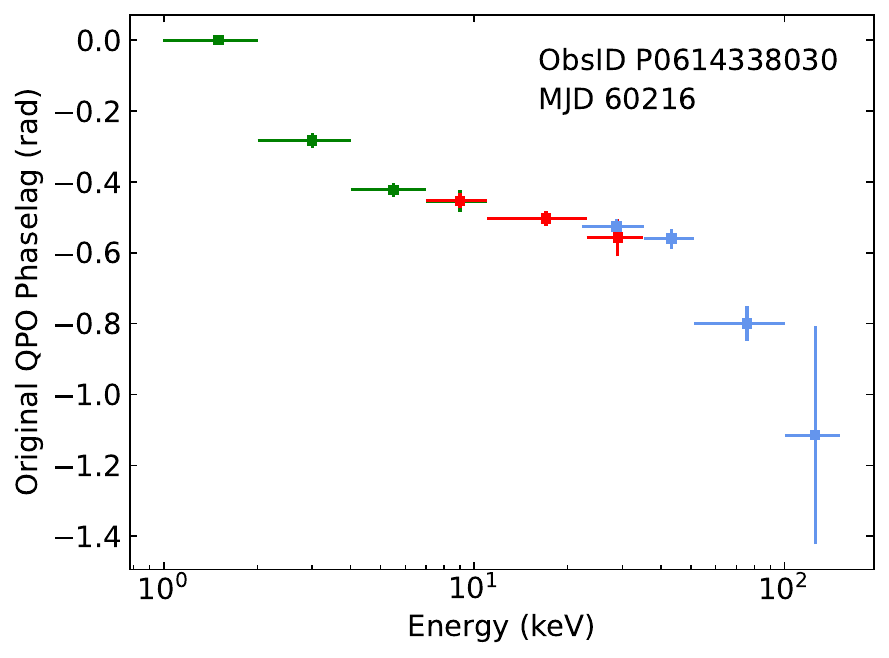}
        \caption{}
        \label{fig:sub3}
    \end{subfigure}
    \hspace{1cm}
    \begin{subfigure}{6cm}
        \centering
        \includegraphics[width=\linewidth]{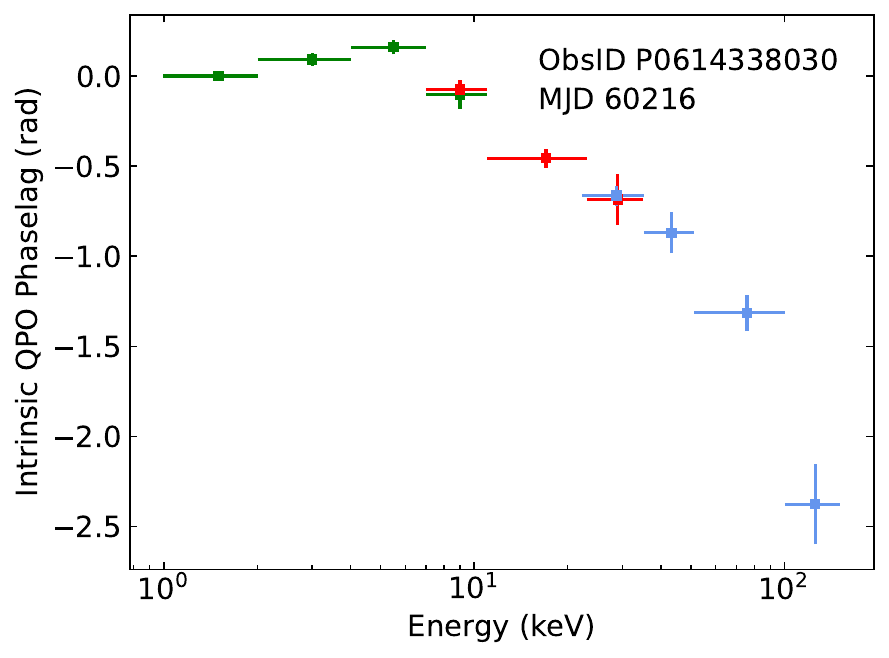}
        \caption{}
        \label{fig:sub4}
    \end{subfigure}
    \caption{Phase lag spectra with a high (\textasciitilde6 Hz) type-C QPO frequency. The reference energy band used to calculate the lag-energy spectra is 1-2 keV. (a) Phase lags as a function of frequency between 1-2 keV and 25-150 keV. (b) Phase lags as a function of frequency between 1-10 keV and 2-10 keV. (c) The "original" LFQPO phase lags as a function of photon energy. (d) The "intrinsic" LFQPO phase lags as a function of photon energy.}
    \label{fig11}
\end{figure*}

We present the \emph{Insight}–HXMT light curves, hardness, and total fractional rms of Swift 1727.8--1613 during the 2023 outburst in Fig.~\ref{fig1}. The LE count rate (2-10 keV) increases rapidly during the rising phase, reaching a peak of 3100 cts/s on MJD 60185, and then slowly decreases to 2500 cts/s on MJD 60197. 
Following this, multiple flares are observed, with peak flux reaching up to 4200 cts/s.
The ME count rate (10-35 keV) rises from 1400 cts/s on MJD 60181 to 1900 cts/s on MJD 60184 and then falls to 350 cts/s on MJD 60222. The HE count rate (25-150 keV) steadily decreases from 3500 cts/s on MJD 60181 to 630 cts/s on MJD 60222, after which it remains at a constant level, similar to the ME count rate. Interestingly, during the period when the flares are detected in the LE band, corresponding dips are observed in the light curves of both the ME and HE energy bands. 

The hardness is defined as the ratio of the count rate in the 4-10 keV energy band to the count rate in the 2-4 keV energy band. In the middle panel of Fig.~\ref{fig1}, the hardness initially declines steeply, then maintains a constant level up until just before the first flare. Following this, the hardness begins to decrease once more, exhibiting an inverse correlation with the LE count rate during the flares.
The evolution of rms is similar to that of hardness. During the flares, the rms markedly decreases, suggesting that the increase in flux is primarily driven by the non-variable component.

The hardness-intensity diagram (HID) and the hardness-rms diagram (HRD) are shown in Fig.~\ref{fig2}. 
The outburst starts at the right part of the plots. As the intensity increases, the source on the HID starts moving to the upper left. The hardness and fractional rms show an approximate linear correlation. For clarity, we categorize the outburst into two distinct states. The initial state, referred to as the `normal state', is similar to the eruptions observed in other typical black holes, which exhibits a segment of the standard q-shaped pattern on the HID. The transition from the `normal state' to the second state occurred near MJD 60198, as indicated by the black dashed line in the figure. This second state is characterized by multiple flares and a gradual decrease in hardness, and is referred to as the `flare state'.

\subsection{Power Density Spectra}

In order to investigate the fast time-variability properties of Swift J1727.8--1613, the PDS is computed for each observation. After subtracting the Poisson noise, the PDS is normalized using the Miyamoto normalization \citep{1991ApJ...383..784M}. Subsequently, the PDS is fitted with a multiple-Lorentzian model \citep{2002ApJ...572..392B}. 
In Fig.~\ref{fig3}, we present six representative PDS of the LE energy band, each with its best fit. The PDS shows two broad noise components (N1 and N2) and one or more peaks (Q1, Q2, and Q3) at varying frequencies. Here, Q1, Q2, and Q3 correspond to the fundamental frequency, the second harmonic of the LFQPO, and the peaked noise components, respectively. 

Type-C LFQPOs were detected in the majority of the observations, while no Type-B or Type-A QPOs were found. The classification of the LFQPOs was based on the criteria proposed by \citet{2005ApJ...629..403C}.  It is worth noting that the peaked noise components are only observed in the `normal state' of the outburst, before the first flare. 

\subsection{Evolution of the PDS}

Next, we study the evolution of the QPOs. Fig.~\ref{fig5} illustrates the evolution of the LFQPO over time. During the observations, the centroid frequency of the LFQPO evolves within a range of 0.1 to 7.8 Hz. 
During the flares, a noticeable increase in the LFQPO frequency is observed. The quality factor of the LFQPOs increases rapidly in the early stages, then maintains a stable level for an extended period, before it starts to decrease. The fractional rms of the LFQPOs shows a rapid increase in the rising phase of the outburst, reaching its maximum value of 14\%, and then progressively declines. Similar to the total fractional rms, the fractional rms of the LFQPOs also exhibit corresponding dips during the flares.

Fig.~\ref{fig6} shows the correlation between the LFQPO frequency and the count rates of LE, ME, HE, as well as the total fractional rms. During the early rising stage and each flare period, the LFQPO frequency exhibits a clear positive correlation with the LE count rates. However, the correlation between the LFQPO frequency and total rms is relatively smooth, with the rms steadily declining as the QPO frequency increases.

Fig.~\ref{fig7} shows the evolution of the peaked noise over time. The evolution patterns of the centroid frequency and quality factor of the peaked noise are similar to those of the LFQPO. However, the evolution of its fractional rms differs. In the early stages of the outburst, the fractional rms of the peaked noise rapidly drops from a maximum of 15\% and stabilizes at a relatively low level.

We also investigated the relations between the frequencies of the peaked noise (Q3) and low-frequency noise (N1) with the frequency of LFQPO (Q1). These relations are shown in Fig.~\ref{fig8}. 
It can be seen that the frequency of peaked noise is correlated with the frequency of LFQPO and follows the PBK relation \citep{1999ApJ...520..262P}. The frequency of LFQPO and low-frequency noise display an approximate piecewise linear correlation. When the frequency of LFQPO is below 5 Hz, it generally agrees with the WK relation \citep{1999ApJ...514..939W}, but deviates when it exceeds 5 Hz. It is worth noting that this break frequency aligns with the frequency at which the Q factor of the LFQPO begins to decrease, as shown in Fig.~\ref{fig5}.

\subsection{Energy dependence of the LFQPO}

In order to quantitatively examine the energy-dependent behavior of the QPO properties, we extract power spectra in several energy bands: LE (1–3 keV, 3-7 keV, 7–11 keV), ME (7–11 keV, 11–23 keV, 23–35 keV), HE (23–35 keV, 35–50 keV, 50–100 keV, 100–150 keV). The fractional rms and the centroid frequency of the LFQPOs as functions of photon energy are shown in Fig.~\ref{fig9}.
We consider the background contribution to the fractional rms calculation. The formula is $rms = \sqrt{P*(S+B)/S}$ \citep{2015ApJ...799....2B}, where S and B stand for source and background count rates respectively, and P is the power normalized according to Miyamoto \citep{1991ApJ...383..784M}. In the region where LE and ME or ME and HE overlap, there is a good agreement between the two detectors.

We discovered that when the frequency of the QPO is low, it scarcely changes with energy. However, as the QPO frequency increases, it exhibits a positive correlation with energy below 10 keV, and remains constant above 10 keV.
The correlation between the QPO rms and the photon energy also displays different patterns depending on the QPO frequency.
When the QPO frequency is low, the rms exhibits an almost logarithmic linear correlation with energy, increasing from 11\% in the lowest energy band to 23\% in the highest energy band. As the QPO frequency increases, a cut-off energy appears in the rms-energy relation, after which the rms decreases with energy. This cut-off energy gradually decreases as the QPO frequency increases, and when the QPO frequency is around 6 Hz, it decreases to around 30 keV.

\subsection{Phase lags of LFQPO}

To study the phase lag of LFQPO and its energy dependence, we calculated the lag-frequency spectra for multiple energy bands, as shown in Fig.~\ref{fig10}. A positive (hard) lag means that the hard photons lag behind the soft ones. We found that, below 10 keV, a distinct hump shape characterizes the phase lag at the LFQPO frequency. However, this hump transforms into a dip in the high-energy band.
It is crucial to clarify that the phase lag directly measured at the QPO frequency using the lag-frequency spectrum does not accurately represent the lag of the QPO itself because of the interference brought from the underlying strong band-noise \citep{2021NatAs...5...94M,2022MNRAS.515.1914Z,2023ApJ...951..130Y}. In order to estimate the intrinsic lag of the Lorentzian associated with the QPO, we employed the methods presented in \citet{2021NatAs...5...94M} and \citet{2022MNRAS.515.1914Z}. This process involved subtracting the average lag of the band-noise, indicated by the red dashed line in Fig.~\ref{fig10}, from the original QPO phase lag in the lag-frequency spectrum.

Fig. \ref{fig10:sub3} and \ref{fig10:sub4} show the QPO phase lag as a function of energy, in which the phase lag directly measured at the LFQPO frequency from the lag-frequency spectrum is referred to as the "original phase lag", and the phase lag after subtracting the contribution from the noise component is referred to as the "intrinsic phase lag". The frequency ranges used for the QPO are chosen to be the Full Width at Half Maximum (FWHM) centered on the centroid for the fitted Lorentzian model in the power spectrum. It is evident that the "intrinsic soft lag" increases with energy below 10 keV, then decreases to a negative value as energy increases, a pattern markedly different from the "original phase lag".

The phase lag of the QPO also undergoes significant evolution with changes in the QPO frequency. Fig.~\ref{fig11} presents the lag spectra when the QPO frequency is above 6 Hz. It is noticeable that the QPO in the low-energy band exhibits a significant negative lag, and the hump at the QPO frequency nearly vanishes. However, in the high-energy band, the dip at the QPO frequency remains. Through the lag-energy spectra, we can find that the original phase lag shows an opposite evolution trend compared to the QPO at low frequency, while the shape of the intrinsic phase lag is similar to that at low QPO frequency. This indicates that the difference in the original lag spectra is mainly contributed by the noise component, thereby validating the reliability of this method of measuring the intrinsic phase lag of QPO.

\section{DISCUSSION}
\label{sec4}

In this work, we have presented the timing results of the new black hole candidate Swift J1727.8--1613 during its 2023 outburst with \emph{Insight}-HXMT observation. The outburst evolution agrees well with the diagrams typically observed in BHTs \citep{2005A&A...440..207B,2011MNRAS.415..292M,2010LNP...794...53B}. Only type-C QPOs are detected during the outburst. The QPO frequency evolves from $\sim$0.1 Hz to $\sim$8 Hz with the increasing X-ray flux (accretion rate). During the initial stage of the outburst, a peaked noise component is detected on the PDS. Additionally, this component exhibits an opposite evolution of fraction rms compared to the LFQPO. This component can only be detected in the `normal state' and becomes completely invisible right before the first flare. 

We investigated the correlations among the low-frequency noise, LFQPO, and peaked noise frequencies. We find that the correlation between the frequencies of the peaked noise and LFQPO aligns well with the PBK relation \citep{1999ApJ...520..262P}. The frequencies of LFQPO and low-frequency noise follows an approximate piece-wise linear correlation. When the frequency of LFQPO is below 5 Hz, this correlation generally agrees with the WK correlation \citep{1999ApJ...514..939W}, while it deviates when the LFQPO frequency exceeds 5 Hz. The PBK and WK correlations have been found in numerous sources \citep[e.g.][]{2000MNRAS.318..361N,2001ApJ...563..239K,2003ApJ...586..419K,2002ApJ...572..392B,2008ApJ...675.1407K, 2015ApJ...799....2B, bu2017low}, implying that different systems might share a common origin of QPOs. The PBK correlation was initially found between the low-frequency kilo-Hertz QPOs and high-frequency kilo-Hertz QPOs in high-luminosity neutron star systems \citep[e.g.][]{1998ApJ...501L..95P}. \citet{1999ApJ...520..262P} and \citet{2000MNRAS.318..361N} later expanded this correlation to the LFQPOs and peaked noise components for low-luminosity neutron stars and BH-LMXBs systems. A recent study found that the high-frequency peaked noise in MAXI J1348--630 also follows the PBK relation with the LFQPO fundamental frequency \citep{2022MNRAS.514.2839A}.

The PBK correlation covers a broad frequency range of three orders of magnitude, with the LFQPOs frequencies ranging between 0.1\ Hz and 100\ Hz \citep{1999ApJ...520..262P}. At the high-frequency end of the PBK correlation, where LFQPO frequencies exceed 10\ Hz, the correlation primarily manifests between the frequencies of LFQPOs and HFQPOs. Subsequent observations also showed that the HFQPOs frequencies typically lie above 50-100\ Hz \citep[e.g.,][]{2012MNRAS.426.1701B, 2019NewAR..8501524I}. At the lower end of the PBK correlation, where LFQPO frequencies are below 10 Hz, the frequencies of HFQPOs are substituted by frequencies of the peaked noise \citep{1999ApJ...520..262P}.  It is reasonable to speculate that a fundamental connection exists between these peaked noises and HFQPOs, particularly in relation to their physical mechanisms.

On the PDS of Swift J1727.8–1613, the Q3 component stands out distinctly. Unlike the broad noise component, it has a smaller FWHM and a higher Q factor. During the initial stages of the outburst, as the source brightness increases, the Q3 component narrows and the Q factor gradually increases, peaking at around 5. At this point, Q3 is more like a QPO than noise. If identifying Q3 with a Q factor larger than 2 as `HFQPOs' (despite their lower frequencies compared to typical HFQPOs), this would be the first evidence of the peaked noise transitioning into a `HFQPO' in BHTs, which would significantly extend the PBK correlation at lower frequencies. As shown in Figure 7b, the PBK correlations found between the LFQPOs and peaked noise components, as well as between the LFQPOs and HFQPOs, are completely consistent in Swift J1727.8-1613. This strongly suggests that the peaked noise components and the HFQPOs probably originate from the same mechanism and would be substituted by the other during the outburst.

In the model proposed by \citet{1998ApJ...492L..53C}, HFQPOs are the result of disc precession induced by the Frame Dragging Effect. When applying this model to several black holes (with HFQPO frequencies ranging from 9 Hz to 300 Hz) with dynamically determined masses, their angular momenta can be derived and the results are in good agreement with those derived from spectral data \citep{1998ApJ...492L..53C}. Assuming that the peaked noises and HFQPOs have the same origin, we apply this model to Swift J1727.8--1613. By adopting the highest frequency of `HFQPOs' found in this system, we made a constraint on the spin of the black hole. For BH mass of 10 and 20 solar masses, the estimated lower limits of the spin are 0.7 and 0.82, respectively. This indicates that Swift J1727.8--1613 may be a high-spin system. A detailed analysis of the peaked noises and HFQPOs, including their energy dependence and connections with flares, will be presented in a forthcoming paper (Yu et al., in preparation).

Subsequently, we investigate the energy dependence of the LFQPO.
We find that the correlation between the 
QPO fractional rms and energy significantly evolves with the QPO frequency (see Fig.~\ref{fig9}). At lower QPO frequencies, the fractional rms of the QPO increases with energy in a logarithmic linear correlation. As the QPO frequency increases, a cutoff energy emerges, beyond which the rms decreases with energy. This cutoff energy gradually decreases as the QPO frequency continues to increase. This is significantly different from what has been observed in other black holes.
\citet{2018ApJ...858...82Y} computed the fractional rms spectrum
of the QPO within the framework of the Lense-Thirring (L-T) precession model \citep{ingram2009low}.
They found that when the system is being viewed at a large inclination angle (i>60), the rms contributed by the reflection component forms an approximate logarithmic linear correlation with energy. Conversely, the rms contributed by the Comptonization component exhibits a cutoff energy. These findings align with the two types of rms-energy correlations we observed in Swift J1727.8--1613. Therefore, we hypothesize that Swift J1727.8--1613 has a high inclination angle, and the evolution of the rms-energy correlation is driven by the distinct contributions from the Comptonization and reflection components.

We also find that the phase lag of LFQPOs shows distinct characteristics in low and high energy bands: it manifests as a hump in the low energy band and transforms into a dip in the high energy band. By adopting the method from \citet{2021NatAs...5...94M} and \citet{2022MNRAS.515.1914Z}, we have measured the intrinsic phase lag of LFQPOs. For different QPO frequencies, the intrinsic phase lag-energy correlation exhibits similar patterns. A similar QPO phase lag--energy relation was also observed in MAXI J1820+070 \citep{2021NatAs...5...94M,2023ApJ...948..116M,2023ApJ...951..130Y}. \citet{2021NatAs...5...94M} explained such phase lag behavior of the QPO by employing a compact jet with precession. In this scenario, the high-energy photons come from the bottom part of the jet closer to the black hole, while the precession of the compact jet gives arise to the QPO and allows the high-energy photons to reach the observer first, resulting in a soft lag. However, this model can only explain soft lags above 10 keV. Recently, they expanded the jet precession model to 0.2-200 keV by assuming that the observed QPOs are generated by the L-T precession of both the jet and the inner disc ring \citep{2023ApJ...948..116M}. In this model, the QPO phase lag in different energy bands due to the hybrid contributions from the jet and the accretion disc. This can naturally explain the phase lag in the low-energy band in Swift J1727.8--1613. Therefore, similar to MAXI J1820+070, the low-frequency QPO in Swift J1727.8--1613 can also be explained by the disc-jet co-precession model.

From the inclination dependence of phase lags
in a sample of 15 black hole binaries, \citet{2017MNRAS.464.2643V} found that the phase lag of the type-C QPOs
strongly depends on the inclination. All samples possess
a slightly hard lag at low QPO frequencies. However, at high frequencies, sources with high inclinations shift to soft lags, while lags in sources with low inclinations become harder. These results substantiate the geometric origin of type-C QPOs. Considering the original QPO phase lags below 20 keV, it is evident that Swift J1727.8--1613 aligns with the trend of high-inclination sources ($i$>70 deg), as outlined in \citet{2017MNRAS.464.2643V}.

In addition to the potential relation between the phase lags and the
inclination of the system, \citet{2017MNRAS.464.2643V} also found that the QPO frequency increases with energy when the frequency exceeds 6–7 Hz for high inclination sources. Below that frequency, the QPO frequency remains stable as the energy increases. In the case of Swift J1727.8--1613, the frequency of type-C QPO stays constant with energy at frequencies below 6 Hz, while it increases with energy at frequencies above 6 Hz. This suggests that Swift J1727.8--1613 has a high inclination, a conclusion that is consistent with the findings from the phase lag analysis.

\section{SUMMARY AND CONCLUSIONS}
\label{sec5}

We have conducted a timing analysis of the new BHC Swift J1727.8--1613 using \emph{Insight}-HXMT observations. The summary of our conclusions is as follows:

(1) The frequencies of the peaked noise and LFQPOs in Swift J1727.8--1613 align well with the PBK relation. Additionally, the unique peaked noise component progressively exhibits HFQPO-like characteristics as the source luminosity increases, potentially marking the first observed transition of peaked noise into a `HFQPO'. This finding suggests a intrinsic connection between the peaked noise and the HFQPOs, which could provide significant improvement to the existing PBK relation at lower frequencies.

(2) Based on the assumption that the peaked noise originates from the precession of the accretion disc, we have established constraints on the spin of this source. For BH mass of 10 and 20 solar masses, the estimated lower limits of the spin are 0.7 and 0.82, respectively, which suggests that Swift J1727.8--1613 could be a high-spin system.

(3) The energy-dependent analysis of the frequency, fractional rms, and phase lag of the LFQPO consistently suggest that Swift J1727.8--1613 is a high inclination system.

\section*{Acknowledgements}

This work made use of the data from the \emph{Insight}-HXMT mission, a project funded by the China National Space Administration (CNSA) and the Chinese Academy of Sciences (CAS), and data and/or software provided by the High Energy Astrophysics Science Archive Research Center (HEASARC), a
service of the Astrophysics Science Division at NASA/GSFC. This work is supported by the National Key RD Program of China (2021YFA0718500) and the National Natural Science Foundation of China (NSFC) under grant Nos. U1838202, 12273030, 11733009, 11673023, U1938102, U2038104, U2031205,
12233002, 12133007, 12333007, the CAS Pioneer Hundred Talent Program (grant No. Y8291130K2) and the Scientific and Technological Innovation Project of IHEP (grant No. Y7515570U1). This work
was partially supported by the International Partnership Program of the CAS (grant No. 113111KYSB20190020).

\section*{Data Availability}
 
The raw data underlying this article are available at
http://hxmten.ihep.ac.cn.



\bibliographystyle{mnras}
\bibliography{example} 





\bsp	
\label{lastpage}
\end{document}